\documentclass[twocolumn,showpacs,a4paper,superscriptaddress,nofootinbib,tightenlines,floats]{revtex4} 
\usepackage{bm}
\usepackage{latexsym}
\usepackage{dcolumn}
\usepackage{amsfonts,amssymb}
\usepackage{graphicx}
\usepackage{psfrag}
\usepackage{amsmath}
\begin{document}

\def\apjl{Astrophys. J. Lett.}
\def\mnras{Mon. Not. Roy. Astron. Soc.}
\def\mnrasl{Mon. Not. Roy. Astron. Soc. Lett.}
\def\physrep{Phys. Rept.}
\def\apjs{Astrophys. J. Suppl.}
\def\aap{Astron. Astrophys.}
\def\araa{Annu. Rev. Astron. and Astrophys.}
\def\aj{Astron. J.}

\title{Scalar field dark energy perturbations and the Integrated Sachs Wolfe effect}   

\author{H.~K.~Jassal}

\email[$^a$]{Email: hkjassal@iisermohali.ac.in}

\affiliation{Indian Institute of Science Education and Research Mohali,
S. A. S. Nagar, P. O. Manauli, Mohali-140306, Punjab, India. \\
}

\begin{abstract}
Dark energy perturbation affects the growth of matter perturbations
even in scenarios with noninteracting dark energy.  
We investigate the Integrated Sachs Wolfe (ISW) effect in various
canonical scalar field models with perturbed dark energy.  
We do this analysis for models belonging to the thawing and freezing
classes.
We show that between these classes there is no clear difference for
the ISW effect.   
We show that on taking perturbations into account, the contribution due
to different models is closer to each other and to the cosmological
constant model as compared to the case of a smooth dark energy.
Therefore  considering dark energy to be homogeneous gives an
overestimate in distinction between different models.
However there are significant difference between contribution to the
angular power spectrum due to different models.

\end{abstract}

\preprint{}

\maketitle
The accelerated expansion of the universe has been confirmed by
various observations, including Supernova type Ia observations
observations of Cosmic Microwave Background and large scale structure
\cite{data}.  
The simplest dark energy model, the cosmological constant $\Lambda$
\cite{ccprob_wein,review3,anjan} has the  fine tuning
problem which lead to proposals for various dark energy models
to explain the current accelerated expansion of the universe.
In these models, the equation of state of dark energy is a
function of time and typically the dark energy component is assumed to
be a homogeneous component of the universe.
Dark energy is described by an ideal fluid or by a scalar field and the
parameters of dark energy models are well constrained by observations. 
However, if the background evolution for different models is the same,
we cannot distinguish between them purely on the basis of distance
measurements.

Since dark energy affects the background evolution of the universe, it
affects the growth of perturbations \cite{ujs,hkj,hkj_antho,chpgas_pert,sph_coll}.
In addition dark energy contributes via its inherent perturbations.
If dark energy is a cosmological constant, then it does not cluster. 
The gravitational field $\Phi$ begins to decay when the cosmological constant
starts to dominate \cite{ujs,hkj,hkj_antho}.  
For the case $w \neq -1$, in the matter dominated epoch, the potential
$\Phi$ remains at a constant value and decays when dark energy contribution becomes
important and this rate of decay  depends on the value of $w$.
For a canonical scalar field, dark energy perturbations are correlated with
the matter perturbations leading to an enhancement in matter perturbations
\cite{ujs}, whereas they are suppressed if dark energy is a barotropic
fluid  compared to the corresponding case of a homogeneous dark energy
model \cite{hkj}. 
The difference is due to the fact that the pressure gradients evolve
differently in these model and the homogeneous limit is arrived at 
differently for the two  classes of models.  
In particular, the evolution of perturbations in different dark energy
models is expected to break the degeneracy between those  which cannot
otherwise be distinguished by distance  measurements 
\cite{weller_lewis,bean_dore,depert,gordonhu}.

Dark energy perturbations affect the low $l$ quadrupole in the CMB
angular power spectrum through the ISW effect\cite{weller_lewis,bean_dore} .
The ISW effect can distinguish a cosmological constant from other
models of dark energy \cite{isw_obs,ddw}.  
The difference in  models is expected to make a
significant contribution to integrated effects such as the ISW effect
\cite{ddw}. 
At scales smaller then the Hubble radius, dark energy can be assumed to be
homogeneous and for these scales, a fluid model is good approximation
to a scalar field model \cite{hkj}.
For large scales the behavior of perturbation growth depends on
specific models and the growth factors deviate from each other (for
scalar tensor gravity models, it has been shown that the dark energy
perturbations are significant even at small scales \cite{scalten_pert}) . 
Since dark energy perturbations are significant at large scales, i.e., scales
larger then Hubble radius, the effect of dark energy perturbations is
expected to show prominently in the Integrated Sachs Wolfe (ISW) effect.

\begin{figure}
\begin{center}
\begin{tabular}{cc}
\includegraphics[width=0.45\textwidth]{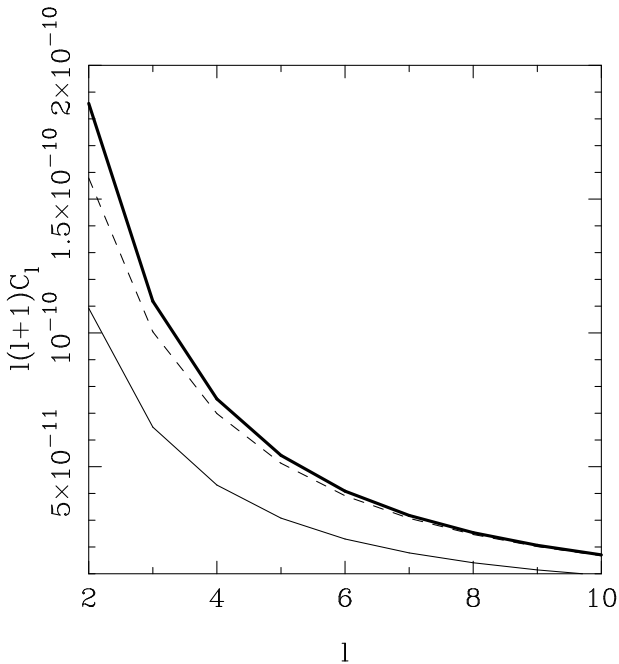} 
\end{tabular}
\end{center}
\caption{This figure shows plot of the ISW contribution to the angular power spectrum for the
  potential the  $V(\phi) = M^{4+n}
\phi^{-n} exp(\alpha \phi^2/M_P^2)$ with$\alpha=0.8$ and $n=4$ and for
the cosmological constant model. The thick solid line corresponds to
a homogeneous scalar field, the dashed line is for perturbed dark
energy and the thin solid line is for cosmological constant model.} 
\label{fig1}
\end{figure}

In this Report, we study the ISW effect in various (canonical) scalar
field models, if the dark energy is considered to be a smooth
component or if it contributes to structure formation by way of its
perturbations. 
As in \cite {ujs,hkj,hkj_antho}, we choose the  Newtonian gauge
\begin{equation}
ds^2 = (1+ 2\Phi) dt^2 -  a^{2}(t) \left[(1- 2\Phi) \delta_{\alpha \beta}
  dx^{\alpha} dx^{\beta}\right] 
\end{equation}
where $\Phi$ is the gauge invariant potential \cite{bardeen}. 
The linearized Einstein equations obtained from  this metric are
\begin{eqnarray} \label{eq::einstein}
\frac{k^2}{a^2} \Phi + 3 \frac{\dot{a}}{a} \left(\dot{\Phi} +
  \frac{\dot{a}}{a} \Phi \right) &=& - 4\pi G  \left[\rho_{NR} \delta_{NR} +
  \rho_{DE} \delta_{DE} \right] \\ \nonumber 
\dot{\Phi} + \frac{\dot{a}}{a} \Phi &=& -4 \pi G \left[\rho_{NR} v_{NR} +
  \rho_{DE} v_{DE}\right] \\
\nonumber 
4 \frac{\dot{a}}{a} \dot{\Phi} + 2  \frac{\ddot{a}}{a} \Phi +
\frac{\dot{a}^2}{a^2} \Phi + \ddot{\Phi} &=& 4 \pi G  \delta P
\end{eqnarray}
where dot is the derivative with respect to  time $t$.
The potential for the matter peculiar velocity is given by $v_{NR}$ with
$\delta u_{i} = \nabla_{i}v_{NR}$. 
The perturbed quantities have been Fourier transformed and we have
replaced $\nabla^{2}$ by $-k^{2}$, where $k$ is the wave number
defined as $k= 2\pi/\lambda$, where $\lambda$ is the comoving length
scale of the  perturbation (We have dropped the subscript $k$ from the
perturbed quantities).

We have assumed a spatially flat cosmology and to solve the equations
we choose the initial equation of state $w \approx -1$ at early times
(say at a initial redshift of $z_{i} = 100$).   
We fine tune the value of the initial value of the scalar field such
that the universe begins to accelerate at late times and the
parameters are such that the matter density parameter is within the
range allowed by the present day observations.  
We need two second order equations which connect $\Phi(t)$ and $\delta
\phi(t)$ and we  choose the third equation in system \ref{eq::einstein}. 
The equation for  $\delta \phi(t)$ (the perturbation in scalar field) is given by \cite{ujs}:  
\begin{equation}
\ddot{\delta \phi} + 3\frac{\dot{a}}{a}\dot{\delta \phi} + \frac{k^2
  \delta \phi}{a^2} + 2\Phi V'(\phi)  -4\dot{\Phi}\dot{\phi} +
V''(\phi)\delta \phi = 0, \label{eqn::pertkg}
\end{equation}
where $V(\phi)$ denotes the scalar field potential.
We assume the initial perturbation in the scalar field to be
negligibly small compared to other perturbed quantities ($\Phi$ and
$\delta_{NR}$) and we can set the initial conditions as $\delta
\phi_{i} = 0$ and $\delta \dot{\phi}_{i} = 0$.  
In the matter dominated epoch, the gravitational potential
$\dot{\Phi}(t)=0$ for all values of $k$.

\begin{figure*}
\begin{center}
\begin{tabular}{cc}
\includegraphics[width=0.45\textwidth]{allmodels.ps} 
\includegraphics[width=0.45\textwidth]{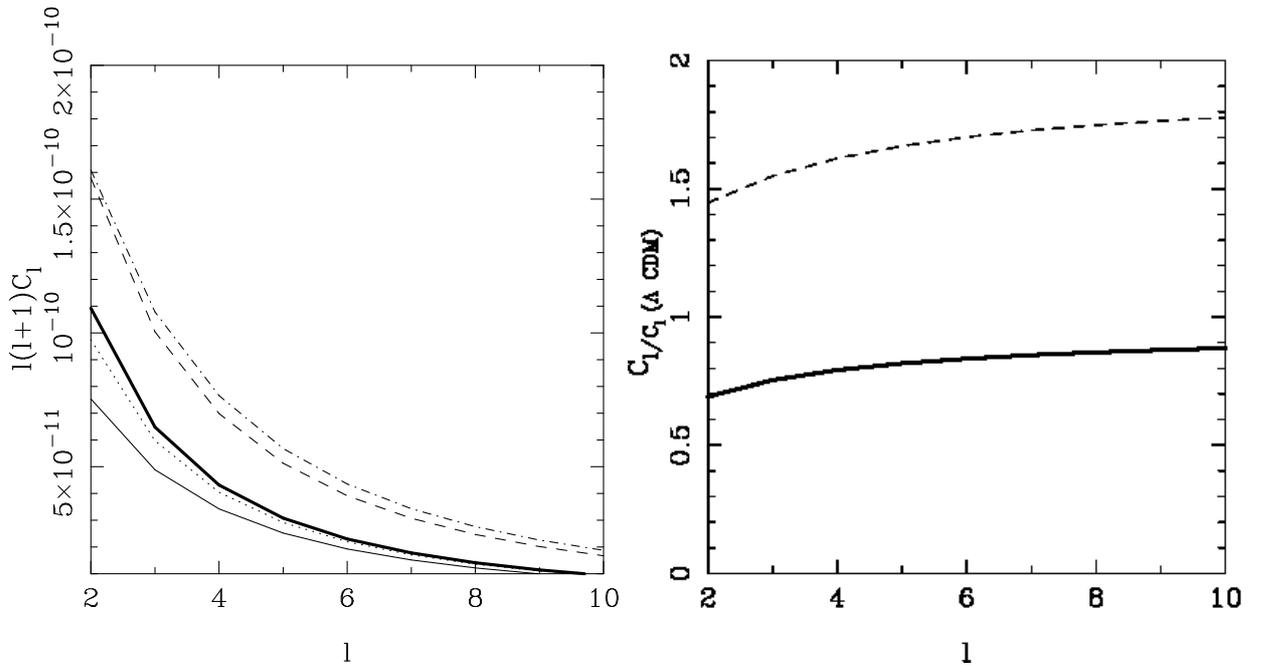} 
\end{tabular}
\end{center}
\caption{In this figure the plot on the left shows the angular power
  spectrum (ISW) for the different models considered in this analysis.  The
  solid line  corresponds to the scalar field  (freezing) potential
  $V(\phi)=M^{4+n} \phi^{-n}$ (F1) where we have taken $n=2$.    The
  dashed line corresponds to   potential $V(\phi)=M^{4+n} \phi^{-n}
  exp(\alpha \phi^2/M_P^2)$ (F2) with   parameters $\alpha=0.8$ and $n=4$. 
The dotted line denotes the thawing exponential potential $V(\phi) = M^4
exp(-\sqrt{\alpha} \phi/M_P)$ (T1) with $\alpha=1$ and the dot-dashed line
corresponds to  $V(\phi)=M^{4-n} \phi^{n}$ (T2) where we have taken
$n=2$. The thick solid line corresponds to the $\Lambda$CDM model. The
line styles corresponding to different 
models are chosen to be same as in \cite{hkj_antho}. The plot on the
right shows the ratios of values of angular    power spectrum for a
freezing and a thawing model. The models are   respectively, model F1
with $\alpha=0.8$ and $n=4$ and model T1 with   the value   $\alpha
=1$.}   
\label{fig2}
\end{figure*}

The ISW temperature fluctuations are due to photons climbing in and
out of potentials which are functions of time \cite{isw_obs}.
The temperature perturbation is given by
$$
\tau = \frac{\Delta T}{T_{CMB}}=-\frac{2}{c^2}\int_{\eta^*}^{\eta_0}  d \eta \dot{\Phi}
$$
and the angular auto correlation  power spectrum is given by
$$
C_l = \frac{2}{\pi} \int k^2 dk P(k) I_l^{2}(k)
$$
where $P(k)\propto k^n T^2(k) $ is the present power spectrum with
$T(k)$ being the transfer function. 
For the transfer function, we adopt the fitting function given by \cite{bardeen2} 

\begin{eqnarray}
T(q \equiv k/(\Omega_{NR}h Mpc^{-1})) =
\frac{ln[1+2.34q]}{2.34q}~~~~~~~~~~~~& &
\\ \nonumber
 \left[1 +
  3.89q +  (16.2q)^2 + (5.47q)^3 + (6.71q)^4 \right]^{-0.25} &&
\end{eqnarray}
and the function $I$ is given by
$$
I_l(k)=-2 \int \frac{d\Phi(z)}{dz} j_l[k\chi(z)]dz
$$
where $j_l$ is the spherical Bessel function and $\chi$ is the
coordinate distance.

We use the classification scheme described in \cite{limitsofq},for the
freezing and thawing models.
The scalar field potentials considered in our analysis are the
'thawing' potentials: the exponential potential \cite{expo}, $V(\phi)
= M^4 exp(-\sqrt{\alpha} \phi/M_P)$ (henceforth,  model T1), 
Polynomial (concave) potential \cite{linear,quadratic}, $V(\phi) =
M^{4-n} \phi^n$ (model T2) and for 'freezing' behavior, the Inverse
power potential \cite{invphi}  $V(\phi) = M^{4+n} \phi^{-n}$ (model
F1) and $V(\phi) = M^{4+n} \phi^{-n} exp(\alpha \phi^2/M_P^2)$
\cite{invexpo} (model F2).   
In the freezing type models, the scalar field  rolls down a steep
potential, remains subdominant and at late times dominates and drives
the acceleration of the expansion of the universe. 
The thawing models have a nearly flat potential, the equation
of state starts at $w=-1$ at early times and deviates from
this value (thaws) at late times.
Therefore, the dark energy equation of state evolves very differently in these
classes of models.

For freezing potential referred to as F1, the scalar field 
rolls down the steep potential at early times and at late times
freezes at the value $w=-1$.  
The epoch of freezing  depends on the fine tuning of
cosmological parameters.
In this model, if the scalar field is frozen before the present time,
the duration of the matter dominated phase is too small and there is
insufficient structure formation \cite{hkj_antho}.   
Therefore, the scalar field needs to freeze in far future. 
In this model, for $n=2$, the dark energy equation of state reaches a
maximum of approximately $-0.7$, and the initial  $\phi$ can
be fine tuned such that at the present epoch $\Omega_{NR} \approx 0.3$. 
For the potential F2, the freezing behavior is achieved earlier and
the matter dominated phase is sufficiently long.
For a given $n$, the present value of the equation of state deviates
from  $w=-1$ as $\alpha$ increases and for a constant $\alpha$, the
value of the equation of state increases with $n$. 
As in all other models, at very early time we assume $w=-1$.
If $\alpha=0.6$ and $n=1$, the value equation of state reaches $
\approx -0.99$ at redshift $\approx 0.4$ and begins to freeze and if
$\alpha=0.8$ and $n=4$, the maximum  value the equation of state
reached  is $w=-0.82$.  
The values of the equation of state are  within the range allowed
by distance observations.
In canonical scalar field models, dark energy perturbations are
correlated with matter perturbations and consequently enhance them
\cite{hkj,hkj_antho}. 
The enhancement in matter perturbations leads to a suppression
in power at large angular scales as compared to the homogeneous
dark energy case. 
Compared to the cosmological constant model, the matter perturbations
are suppressed and hence ISW effect is stronger. 
The deviation from the cosmological constant case decreases with an
increasing $l$. 
We illustrate the above discussion in  Fig. 1 where we plot the
angular power spectrum for the last model with and without
perturbations and compare with those in the $\Lambda$CDM model.

For thawing  potential T1, larger the value of $\alpha$, more is the
deviation from cosmological constant type behavior and the
perturbations contribute more to power at small $l$ as compared to the
$\Lambda$CDM model. 
For this potential \cite{ujs,hkj,hkj_antho}, if $\alpha=1$, the
present day equation of state is 
$w=-0.86$ and at redshift $z=0.81$, the acceleration of the universe
starts. 
For $\alpha =0.1$, the present day value of equation of state is $w=-0.95$.  
For smaller $\alpha$ where the present equation of state is closer to
$-1$, there is no significant difference if  we assume dark 
energy to be a smooth component  or if we assume dark energy to be a
perturbed component.
If $\alpha=1$ there is a large enhancement of perturbations in matter
\cite{ujs} and this in turn translates to difference in the power at
small $l$ as compared to the smooth dark energy case and that of
cosmological constant. 
For the concave potential (T2) with $n=1$ and $n=2$, the present day
of equation of state is $w \approx -0.94$ if we choose  $n=1$ and
reaches  $w \approx -0.3$ (which is disfavored by observations) for
$n=2$.  
For a quadratic potential, the gravitational potential initially decays,
and in the future the equation of state starts to oscillate between
$w=-1$ and $w=1$. 
At large scales, therefore, there is a large enhancement in
matter perturbations as compared to the case when we assume a smooth dark energy \cite{hkj_antho}. 
In contrast, if $n=1$, the gravitational potential $\Phi$ continues to
decay and does not show ans oscillatory behavior.
Matter perturbations in the linear potential remain close to those
in cosmological constant model and therefore the contribution of the
ISW effect to the angular power spectrum remains close. 
In quadratic scalar field potential case, these perturbations are enhanced
compared to the $\Lambda$CDM model at early times and are suppressed at late
times.
The contribution to the angular power spectrum due to the ISW effect
is therefore the largest including that of freezing potentials.
The above results are summarized in Fig. 2.

We have shown the effect of dark energy perturbations on the
Integrated Sachs Wolfe effect.
We have chosen freezing/thawing type models for this analysis.
One expects the freezing type models to have a
higher rate of growth of density contrast at early times, since the
equation of state of dark energy is further away from that of a
cosmological constant. 
For thawing type models, dark energy perturbations affect the
matter perturbations at late times.
The matter perturbations remain close to those in cosmological
constant model and at small redshifts being to deviate as the equation
of state goes away from $w=-1$. 
The density contrast evolution is significantly different for
different models and from that of the cosmological constant model.
This difference translates into the difference in the Integrated Sachs Wolfe
effect.  
We show that inclusion of perturbations reduces the difference in the
evolution of matter density contrast for different models, hence
including perturbations increases the degeneracy between different
canonical scalar field models and the freezing/thawing classification scheme is no
longer valid.
The differences between various models, however, are still significant and more
observations and cross correlation of ISW effect with large scale
structure indicators may provide a way to distinguish between various
dark energy models.

\end{document}